\documentclass[a4paper]{article}
\usepackage[square,numbers,sort&compress]{natbib}
\usepackage[margin=1in]{geometry}
\usepackage[usenames,dvipsnames]{xcolor}
\usepackage{soul}
\usepackage{enumitem}
\usepackage{ulem}
\usepackage{verbatim}
\usepackage{url}
\usepackage[T1]{fontenc}
\usepackage{ae}
\usepackage[utf8]{inputenc}
\usepackage{amsmath, amssymb}
\usepackage{siunitx}
\usepackage{booktabs}
\usepackage{subfigure}
\usepackage{xspace} 
\usepackage{multicol}
\usepackage{finesse}
\usepackage{fancyvrb}
\usepackage{caption}
\usepackage{wasysym}
\DefineVerbatimEnvironment{finesse}{Verbatim} 
{formatcom=\small}
\usepackage{amsfonts}
\usepackage{mathrsfs}
\usepackage{parskip}
\usepackage{graphicx,float}
\usepackage[english]{babel}
\usepackage{tabularx}
\usepackage{textcomp}
\usepackage{bm}
\usepackage{duckuments}
\usepackage{matlab-prettifier}  
\usepackage{pythonhighlight}    
\usepackage{adjustbox}
\usepackage{hyperref}
\definecolor{linkcolor}{rgb}{.8,0,0}
\definecolor{urlcolor}{rgb}{0,0,.7}
\definecolor{citecolor}{rgb}{0,.5,0}
\definecolor{acrocolor}{rgb}{0,0,.7}
\hypersetup{bookmarksopen,colorlinks=true}
\hypersetup{pdfstartview=FitH}
\hypersetup{linktocpage=true,bookmarksnumbered=true}
\hypersetup{plainpages=false,breaklinks=true}
\hypersetup{linkcolor=linkcolor,citecolor=citecolor,urlcolor=urlcolor}
\definecolor{myblue}{RGB}{0,0,255}
\colorlet{trb}{myblue!50!white} %

\definecolor{carmine}{RGB}{150,0,24}

\newcommand\tick{\bgroup\markoverwith{\textcolor{red}{\rule[0.5ex]{2pt}{0.4pt}}}\ULon}

\title{Mirror Surface Evaluation for the Einstein Telescope\\ Using Virtual Mirror Maps}

\author{
A. Bianchi$^{1,2,*}$,
A. C. Green$^{2,3}$,
J. Degallaix$^{5}$,
F. A. Feldmann$^{2,4}$,
A. Soflau$^{1,2}$,
A. Freise$^{1,2,\dagger}$
}
\date{} 

\begin{document}
\maketitle
\vspace{-0.5cm}
\begin{center}
\textit{
$^1$ Vrije Universiteit Amsterdam, Amsterdam, The Netherlands \\
$^2$ Nikhef, National Institute for Subatomic Physics, Amsterdam, The Netherlands \\
$^3$ Maastricht University, Maastricht, The Netherlands \\
$^4$ Delft University of Technology (TU Delft), Delft, The Netherlands \\
$^5$ Laboratoire des Matériaux Avancés (LMA), Université de Lyon, CNRS/IN2P3, 69100 Villeurbanne, France
}
\end{center}
\begin{center}
    March 24, 2026
\end{center}

\begingroup
\renewcommand\thefootnote{}
\footnotetext{\normalfont $^*$ \href{mailto:a.bianchi@email.com}{a.bianchi@email.com}}
\footnotetext{\normalfont $^\dagger$ \href{mailto:a.freise@email.com}{a.freise@email.com}}
\endgroup


\section*{Abstract}
The performance of mirrors in optical interferometers is critically influenced by their surface quality. Accurate metrology enables mirror surfaces to be characterized through phase maps describing their three-dimensional structure after coating. In this work, we combine Zernike polynomial decomposition and spatial frequency (PSD) analysis with numerical optical simulations to quantify the impact of surface distortions on the reflected optical field. The method is validated using metrology data from mirrors currently installed in the Advanced Virgo gravitational-wave detector. Building on this validation, we introduce a framework for generating realistic virtual mirror maps that reproduce both low order aberrations and high spatial frequency content of measured surfaces. These virtual maps are used in optical simulations to systematically explore and compare candidate surface quality specifications for future detectors, with particular focus on the Einstein Telescope. Our results show that metrology-informed virtual mirrors provide a practical design tool to assess the impact of different surface specifications on optical performance, and to relate future requirements to the performance of existing interferometers.

\section*{Introduction} 
The optical performance of gravitational-wave interferometers is highly sensitive to mirror surface quality: figure errors and coating inhomogeneities scatter light out of the intended beam shape, adding excess noise and degrading interferometer control and sensitivity~\cite{Brooks_16,Rocchi_2012}. 
For existing detectors such as Advanced Virgo Plus (AdVirgo+)~\cite{Acernese_2015} and Advanced LIGO~\cite{Ad_Ligo_2015}, accurate mirror surface metrology, delivered as high-resolution mirror phase maps, has become a standard tool to support optical design, tolerance studies, commissioning, and noise budget analyses. \\
Historically, mirror surface quality specifications for first and second generation gravitational wave
detectors were defined before detailed surface metrology was available. For LIGO and Virgo, early
requirements were therefore based on analytical models of optical cavities and scattering, together with
simplified assumptions on surface distortions and conservative safety margins
\cite{kogelnik_li_1966,hello_vinet_1990}. As a result, surface quality was typically
specified through global figures of merit, such as RMS flatness or limits on low-order aberrations,
rather than full surface maps.
With the advent of advanced detectors, detailed mirror metrology became available and was increasingly incorporated into optical simulations and commissioning studies. This led to the development of Virtual Mirror Maps (VMMs), i.e. digital representations of synthetic mirror surfaces, for example based on measured two-dimensional power spectral density~\cite{Straniero_15}; these maps have long been implemented in tools such as SIS~\cite{Yamamoto2013SIS}.
However, this approach has not been applied to future interferometers whose mirrors are yet to be fabricated, such as the Einstein Telescope (ET)~\cite{ETDesignReportUpdate2020}. In this context, our virtual mirror maps provide a practical tool to bridge this gap by enabling optical simulations based on synthetic,
yet metrology-informed, mirror surfaces. This allows candidate mirror surface quality specifications to
be systematically tested, including whether optics with quality comparable to those of current
detectors would be sufficient to meet the more stringent requirements of ET.\\
In this study, we generate, preprocess, and test several types of virtual mirror maps with statistical properties comparable to those of mirrors currently installed in Advanced Virgo. These maps are constructed to reproduce both the spectral and structural characteristics of measured surfaces and are used as inputs to optical simulations of the ET high-frequency interferometer, allowing candidate surface characteristics to be evaluated prior to fabrication. 
This strategy enables statistical analyses to explore variability across multiple realizations and supports extrapolation to new parameter spaces, such as larger mirror diameters or different spatial distortion structures. Overall, the use of virtual mirror maps provides a practical and quantitative framework for the design phase of the ET interferometers, preserving physical realism while addressing stringent optical performance requirements.
Our study focuses on the impact of spatial defects on the main laser field, i.e.\ the scattering of light from the intended Gaussian beam shape into higher-order optical modes that may continue to propagate in the interferometer \cite{kogelnik_li_1966,BayerHelms1984,Bond2013}.
The lowest order distortions, piston and tilt, are treated separately, as they are actively sensed and corrected by the interferometer alignment and length control systems~\cite{abbott2010LSC}. Whereas curvature distortions are handled independently, as they are compensated by the Thermal Compensation System (TCS), which is designed to sense and mitigate wavefront curvature variations induced by thermal effects in the core optics~\cite{Brooks_16}.\\
We thus compare two methods for preparing mirror surface data for use in optical simulations by removing these features.
By evaluating the resulting optical performance of AdVirgo+ mirrors, we find that a preprocessing based on minimizing the coupling coefficients to Hermite--Gauss optical modes is more effective than to use a Zernike polynomial subtraction. 
The virtual mirror maps are generated using complementary approaches that target different spatial characteristics of measured mirror surfaces. In particular, we construct synthetic maps based on Zernike polynomial reconstruction, spatial frequency synthesis from measured power spectral densities, and a mixed approach that combines low order Zernike structure with high-frequency content reproduced in the Fourier domain.\\
We first validate our methods  by generating VMMs with similar properties to AdVirgo+ optics, confirming that the generated maps can achieve realistic optical and spatial features when compared with measured data. 
The framework is then applied to the mirror design for the ET high frequency interferometer, where tighter requirements and different operating conditions make realistic surface modeling especially critical for achieving the target strain sensitivity~\cite{Hild2008Pushing}. 
The resulting analysis provides guidelines for tuning the generation of VMMs to target specific figures of merit, such as spatial frequency content and carrier beam power loss, while preserving physical realism.
The paper is structured as follows: in Section~\ref{sec:theory_} we present a brief introduction and motivation for this study, the background of the mathematical tools that we use, and the method for generating the virtual mirror maps. We also describe how existing surface maps can be scaled to match the larger mirror sizes designed for the Einstein Telescope. In Section~\ref{sec:map_preparation} we present and compare two preprocessing methods for removing low-order mirror distortions from measured and generate mirror maps. Finally, in Section~\ref{sec:results}, we present the optical performance of the preprocessed virtual mirror maps in the defined AdVirgo+ and ET setups, displaying their statistical and quantitative properties and draw our conclusions.


\section{Generating Virtual Mirror Maps
}\label{sec:theory_}
The main focus of this work is to generate mirrors with controlled randomized surface patterns. To do so we generate virtual mirror maps relying on three elements: measured mirror phase maps (here, from AdVirgo+'s arm cavity optics); Zernike polynomials \cite{zn_bornwolf}, and Fast Fourier Transform~\cite{fft_paper} algorithms. The measured maps serve as the starting point for generating new VMMs, while the mathematical tools have been selected for their ability to capture different features of the mirror surface.  In addition, we provide a semi-deterministic modeling approach in which low-frequency distortions are not completely randomized but are included as characteristic features of the mirror. In the example presented here, the method relies on information provided by the manufacturer regarding the coating procedure; however, the approach is general and can also be applied whenever the type of mirror under investigation exhibit similar surface characteristics.
\subsection{Mathematical framework}
\label{subsec:framework}


\paragraph{Measured phase maps:}
The surface profile of an ideal mirror would be perfectly smooth; however this is impossible to achieve in practice. 
To quantify the surface deviation between the ideal and real surface, the latter is measured using instruments like Fizeau interferometers, profilometers, or Atomic Force Microscopes (AFMs) \cite{Corbella2024}. 
These instruments provide a high-resolution digital representation of the surface topography and are called \textit{phase maps} (distinguished, for example, from reflectivity or birefringence maps). 
Each element of these 2D matrices corresponds to the local height of the mirror surface, providing a precise measurement of the deviation between the real surface and the ideal one.
In this work, we use measured data from mirrors installed in the Virgo detector, with surface measurements, after coating, performed by the Laboratoire des Mat\'{e}riaux Avanc\'{e}s (LMA) in Lyon, France~\cite{Degallaix2019_JOSAA36C85}.

\paragraph{Zernike polynomials:}
The \textit{Zernike polynomials}, $\text{Z}_n^m(\rho,\theta)$, form a set of orthogonal and continuous functions on a unit disk \cite{Zernike1934a}. This basis is commonly used to describe wavefront distortions and large optical surface aberrations, so can describe the low spatial-frequency content of a mirror surface well. Orthogonality ensures individual elements can be combined without affecting each other, even when summed to infinity. We can describe the mirror surface in terms of $\text{Z}_n^m(\rho,\theta)$ elements as:
\begin{equation}\label{eq:M_Zn} 
\text{M}(\rho,\theta) = \sum_{n,m} c_{n,m} \text{Z}_n^m(\rho,\theta) \,\,\,,
\end{equation} 
where $\text{M}(\rho,\theta)$ is the numerical array of the mirror map in polar coordinates $\rho,\theta$; the indices $n$ and $m$ correspond respectively to the radial degree and the azimuthal order of the polynomial. Only combinations satisfying $|m|\le n$ and an even difference $n-|m|$ are allowed, ensuring orthogonality over the unit disk.
. The coefficients $c_{n,m}$ represent the amplitude of each Zernike polynomial contributing to the surface reconstruction, defined and normalized as in \cite{tao2020}:
\begin{equation}\label{eq:c_nm}
c_{n,m} =
\frac{ \sum_{\rho,\theta} \text{Z}_n^m(\rho,\theta) \, \text{M}(\rho,\theta)}
     {\sum_{\rho,\theta} \left(\text{Z}_n^m(\rho,\theta) \right)^2}  \,\,\,.
\end{equation}
This normalization allows an intuitive interpretation of $c_{n,m}$, evaluated in the same units as the provided map data.

\begin{figure}[ht]
    \centering
    \includegraphics[width=0.95\textwidth]{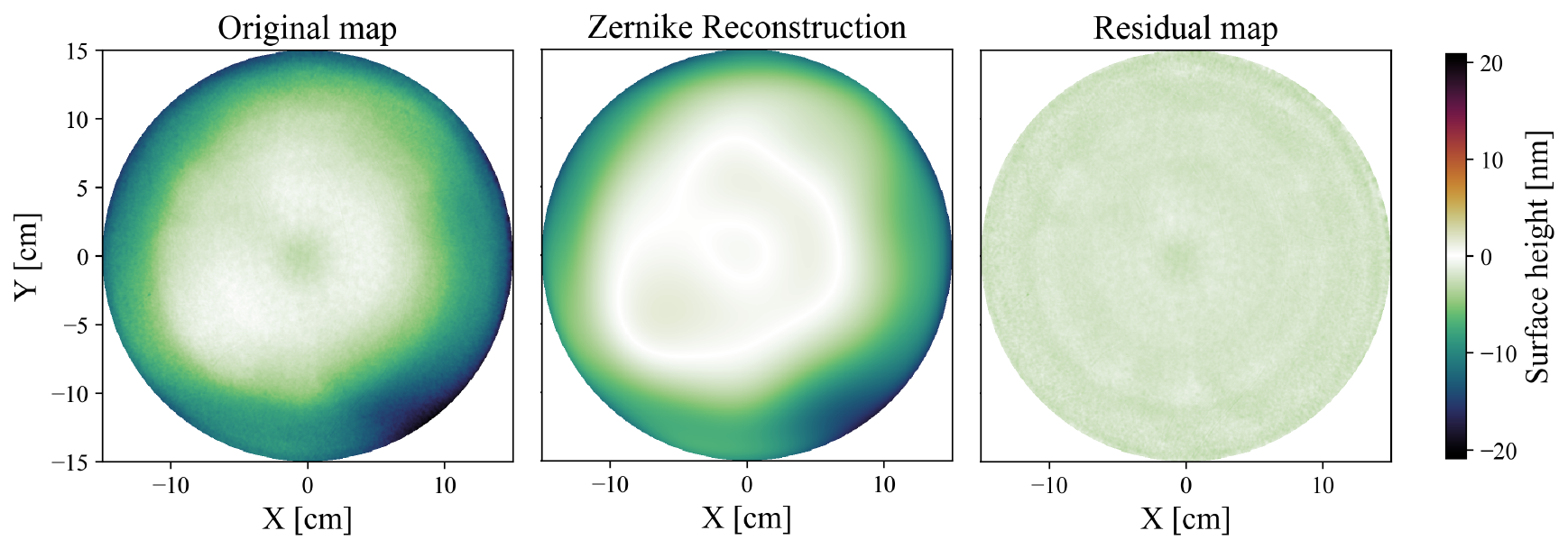}
    \caption{Side by side comparison between a phase map (left), its Zernike reconstruction up to $n=12$, and their residual map (right).}
    \label{fig: prepr_Znrec_res}
\end{figure}

Using Equation~(\ref{eq:M_Zn}) we can reproduce a mirror surface up to a precision that depends on the order, $n$. Figure~\ref{fig: prepr_Znrec_res} compares an original phase map to its reconstruction using $\text{Z}_n^m(\rho,\theta)$ elements up to $n=12$, and plots the corresponding residual map obtained from their difference. This highlights both the capabilities and the limitations of the method: while the Zernike polynomials accurately reproduce large-scale imperfections and capture low-spatial-frequency deformations even at relatively low orders, they fail to represent the surface flatness. 

\paragraph{The Spectral Analysis:}
The \textit{Fast Fourier Transform} (FFT) \cite{fft_paper} is a mathematical algorithm that can efficiently convert spatial data into its frequency domain representation. This enables the analysis of mirror-surface textures, revealing periodic structures and flatness across different spatial frequency scales. Applying the FFT to a mirror map, it is possible to derive its Power Spectral Density in two dimensions (2D PSD) as:
\begin{equation}\label{eq:fft}
\text{2D\,PSD}= \frac{|\text{2D\,FFT}|^2}{f_{\text{x}} f_{\text{y}} \text{N}_{\text{x}} \text{N}_{\text{y}}} \,\,\,,
\end{equation}
where $\text{f}_{\text{x}}$, $\text{f}_{\text{y}}$ are the sampling frequencies and $\text{N}_{\text{x}}$, $\text{N}_{\text{y}}$ the number of points along each axis.
In practice, we  compute the \textit{single-sided} of the $\text{2D\,PSD}$ in two dimensions using the real-input fast Fourier transform 
along the x-direction, and the result is symmetrized along the y-axis using a centered shift 
to obtain a frequency spectrum with positive frequencies only along x and centered frequencies along y. 
The $\text{2D\,PSD}$ matrix is then computed according to Equation~(\ref{eq:fft}). 

The $\text{2D\,PSD}$ is reduced to a one-dimensional spectrum by radially averaging the spectral power over concentric rings in the spatial-frequency plane. Each ring is defined by the radial frequency $f_{\mathrm{R}}=\sqrt{f_x^2+f_y^2}$, and the $\text{1D\,PSD}$ is obtained as the average of the $\text{2D\,PSD}$ within each ring. This effectively reduces the 2D spectral information into a radial profile, assuming isotropy in the surface features (see left image in Figure~\ref{fig:PSD_fftMet}). 
Throughout the paper we plot the one dimension Amplitude Spectral Density (1D ASD) (given $\mathrm{ASD}=\sqrt{\mathrm{PSD}}$, with units $[\mathrm{m}/\sqrt{(1/\mathrm{m})}]$), instead of the 1D PSD, since it conveys the same spectral information while directly reflecting the physical amplitude of the surface height perturbations, making it easier to interpret and compare. 

\begin{figure}[ht]
    \centering
    \includegraphics[width=0.8\textwidth]{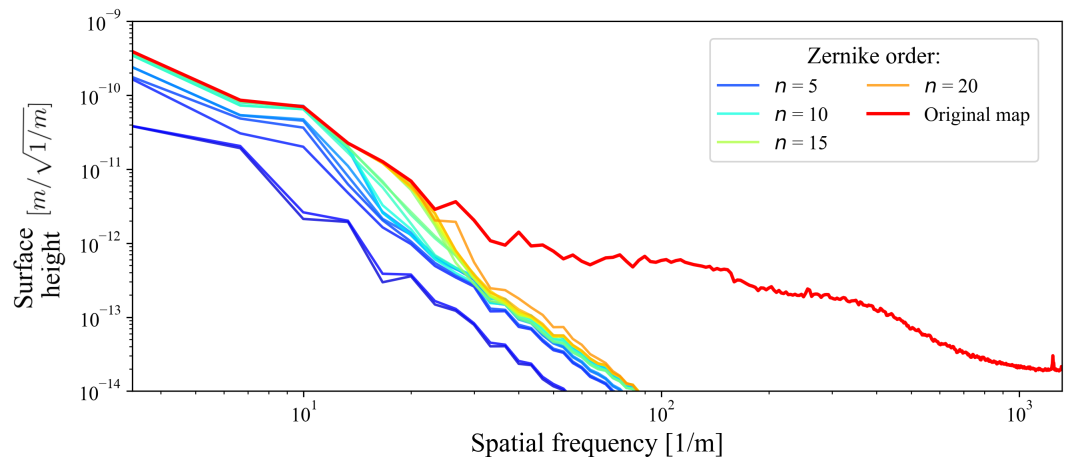}
    \caption{Amplitude Spectral Density of the original mirror map compared with its Zernike reconstructions up to different maximum radial orders $n$. Only a subset of Zernike orders is shown in the legend for clarity. The plot highlights how the different basis orders progressively captures low spatial frequencies as the order increases, but consistently fails to reproduce the high frequency content of the surface.}
    \label{fig:PSD_ZnRec20}
\end{figure}

Several examples of 1D ASDs are shown in Figure~\ref{fig:PSD_ZnRec20}, comparing the original map data with its Zernike reconstructions using polynomials up to $n=20$. Note that the spatial-frequency domain is discrete due to the map sampling; however, for visual clarity, the ASDs are mostly plotted as continuous lines throughout the paper.
The figure  further illustrates why the Zernike polynomials alone are not ideal to describe the full spatial spectrum. Increasing $n$ reduces the mismatch between the original and reconstructed surface, but the convergence toward the full 1D ASD is slow, especially in the high-frequency range. 
Therefore, we use also the Fast Fourier Transform algorithm in our study, to be able to produce Virtual Mirror Maps accountable even for the high frequency domain.

\subsection{Virtual Mirror Map generation}
\label{sec:VMMs}
We developed three distinct randomization methods to generate VMMs, referred to as the \textit{Zernike}, \textit{FFT}, and \textit{mixed} methods. Each includes normalization and randomization schemes, along with specific parameter choices that may be used to tune the final implementation. 

\subsubsection{Controlled randomization methods}
\paragraph{The Zernike method}
\label{method: vmm_zn_method}
We compute the Zernike coefficients of the measured data as outlined above, then generate virtual maps $\text{M}^{\text{Zn}}$ from these coefficients as described in~\cite{tao2020}. We truncate the Zernike expansion at a maximum radial order $n=12$, which corresponds to $91$ polynomials in total.
This choice provides a stable representation of the dominant surface aberrations relevant for optical performance, while avoiding the inclusion of high-spatial-frequency features that are beyond the intended scope of the Zernike representation. The corresponding coefficients $c_{n,m}$ are given by~\eqref{eq:c_nm}, and the ``layer coefficients'' $\text{B}_n$ are defined as:

\begin{equation} 
\text{B}_n = \sqrt{\sum_m (c_{n,m})^2}\,\,. 
\end{equation}
such that new randomized amplitude coefficients $c'_{n,m}$, may be generated while preserving $\text{B}_n$ according to:
\begin{equation}
    c_{n,m}'=\mathrm{B_n} \frac{r_{n,m}}{\sqrt{\sum_{m}(r_{n,m})^2}}\,,
\end{equation}
where the $r_{n,m}$ are randomly drawn from a uniform distribution in the range $[-0.5,0.5]$. The VMM generated using these new random coefficients is computed as in equation~(\ref{eq:M_Zn}), i.e.: 
\begin{equation}
   \text{M}^{\text{Zn}} = \sum_{n,m} c_{n,m}'\text{Z}_n^m\,\,\,.
\end{equation}

When generating ensembles of virtual mirror maps using randomized Zernike coefficients, we introduce an optional global scaling factor \(g=1.9\) applied to the surface height. This factor is not needed to match the PSD of a measured surface when using a deterministic Zernike reconstruction, which by definition reproduces the low-spatial-frequency content of the mirror \cite{CharlotteThesis}. Rather, \(g\) compensates for limitations of the specific randomization scheme adopted here, based on the layer coefficients $\mathrm{B_n}$, which typically yields a lower range of PSD amplitudes than observed in measured maps.

A second motivation for including \(g\) is to obtain carrier power losses in optical simulations that are comparable to those of measured maps. Because Zernike-only maps lack high-spatial-frequency components responsible for wide-angle scattering, matching the total loss of measured mirrors requires enhancing the low-spatial-frequency content, which may modestly exaggerate couplings into low-order higher-order modes and should be borne in mind when interpreting the results.

\paragraph{The FFT method} \label{method: vmm_fft_method}
This algorithm is designed to closely match the 1D PSD of the original data, particularly in the high-frequency tail, such that generated VMMs replicate realistic surface flatness.
The 1D PSD is calculated as described above,  producing a profile representing spectral power as a function of spatial frequency magnitude. 
We generate a new 2D PSD by rotating the 1D PSD around it's origin. We then apply a random phase to each spatial frequency. Finally, by applying the inverse two-dimensional Fourier transform
, a new spatial map is obtained that has different detailed features while preserving the original spectral characteristics. 

\begin{figure} [h]
    \centering
    \includegraphics[width=0.9\textwidth]{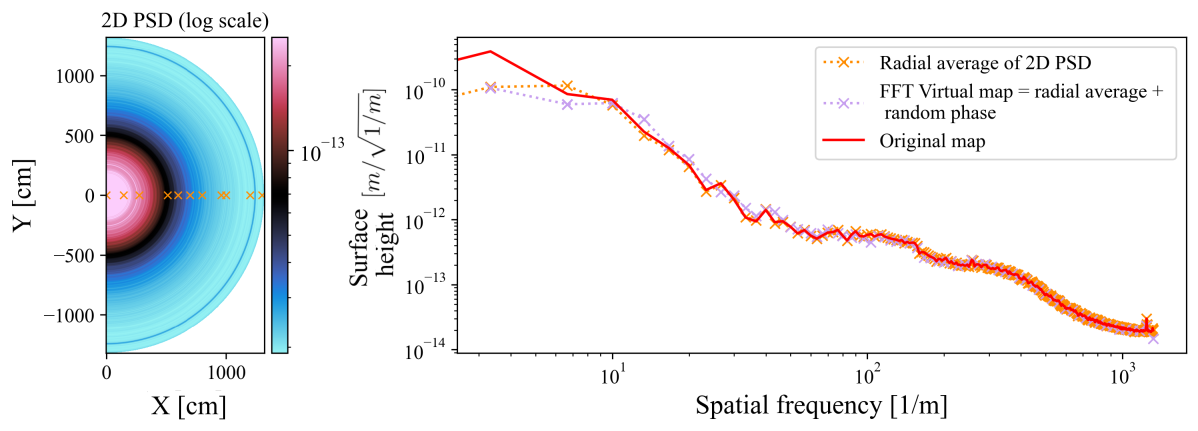}
    \caption{
    Left: Half 2D PSD of the original data, displayed on a logarithmic color scale. The orange crosses on the x-axis correspond to the spatial frequencies used in the radial average of the 2D PSD (values have been plotted downsampled for clarity). 
    Right: 1D ASDs comparison showing the original map, the radial average of the 2D ASD, and the FFT virtual mirror map. The FFT method accurately reproduces the high-frequency tail of the spectrum, but deviates at low frequencies, where the large-scale features of the original surface are not faithfully recovered.}
    \label{fig:PSD_fftMet}
\end{figure}

Figure~\ref{fig:PSD_fftMet} compares the 1D ASDs of the original data and the final virtual mirror maps.  
This illustrates that the FFT based method can generate maps that closely follow the original data 1D ASD, especially in the high-frequency region. However, the method is less effective at low spatial-frequencies, where the amplitudes associated with large-scale or low-order surface features are not faithfully reproduced. 
\paragraph{The mixed method} 
We have developed a new approach for generating virtual mirror maps leveraging the strengths of both the Zernike and FFT methods at low and high spatial-frequencies respectively.

The procedure is as follows. Starting from a measured map, we first generate a  VMM, $\text{M}^{\text{Virt}}_{\text{Zn}}$, using the Zernike method up to order $n$. Next, we generate a second VMM, $\text{M}^{\text{Virt}}_{\text{FFT}}$, using the FFT method. 
To combine the two, we must remove the existing low-spatial-frequency content from the second VMM and replace it with the first VMM. 
To do this, we generate $\text{M}^{\text{Rec}}_{\text{Zn}}$, the $n$th-order Zernike reconstruction of $\text{M}^{\text{Virt}}_{\text{FFT}}$, and subtract it to find the high-spatial-frequency part of the second VMM: 
\begin{equation}
    \text{M}^{\text{Virt}}_{\text{HL}}= \text{M}_{\text{FFT}}^{\text{Virt}} - \text{M}_{\text{Zn}}^{\text{Rec}} \,\,.
\end{equation}
Finally, $\text{M}^{\text{Mix}}$ is created by combining the high and low spatial frequency domains:
\begin{equation}
    \text{M}^{\text{Mix}}=\text{M}_{\text{Zn}}^{\text{Virt}} + \text{M}^{\text{Virt}}_{\text{HL}}\,\,.
\end{equation}

\begin{figure} [h]
    \centering
    \includegraphics[width=0.8\textwidth]{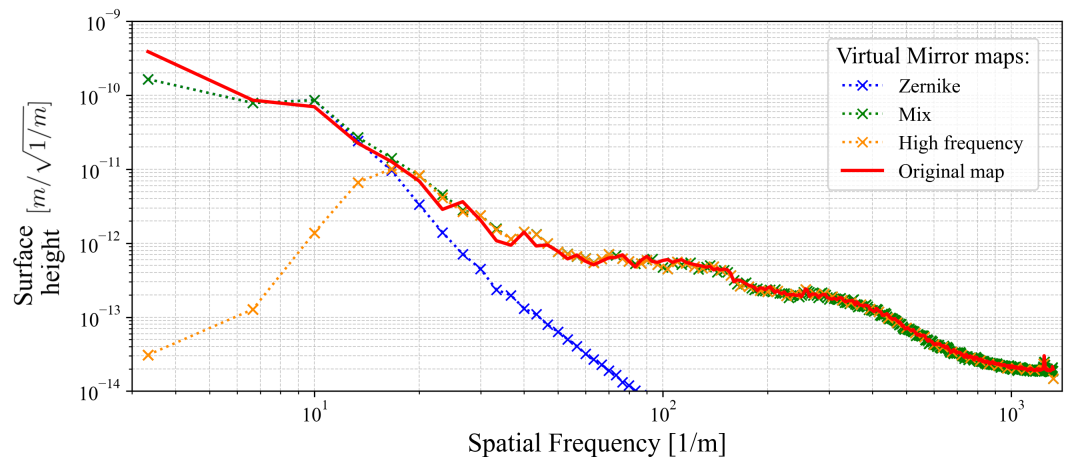}
    \caption{Comparison of the 1D ASDs for the original surface map, the virtual mirror map generated using the Zernike, the high frequency (HF) residual component, and the Mix methods. The plot shows that the mixed approach successfully preserves both the low frequency structure and the high frequency flatness of the original surface. The high frequency spectrum is filtered to reduce point density and to clearly show that both the mixed and HF components closely overlap with the original data.}
    \label{fig:PSD_mix}
\end{figure}

\begin{figure} [h]
    \centering
    \includegraphics[trim=1cm 2mm 1.5cm 1cm clip, width=0.8\textwidth]{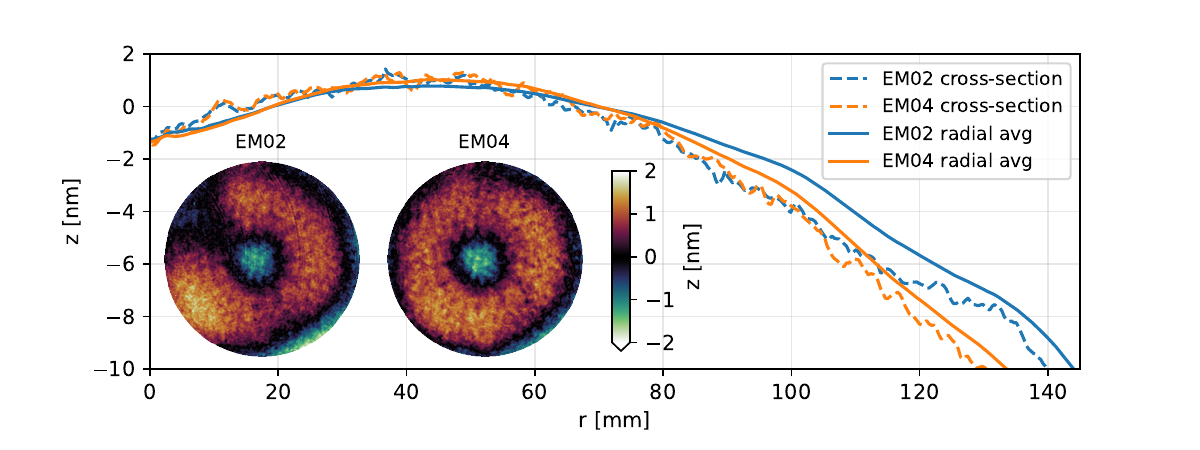}
    \caption{Surface maps of mirrors EM02 and EM04 and their half cross sections after removal of curvature and tilt over the beam area. The insets show only a central 15 cm diameter region for clarity. The similar height distribution highlights the good flatness in the central zone.}
    \label{fig:EMs_coated_zoom}
\end{figure}

As shown in Figure \ref{fig:PSD_mix}, the 1D ASD of the mixed method closely follows the original map data across the entire frequency range, validating the strategy of splitting and recombining spatial frequency domains to generate realistic synthetic mirror maps.
\paragraph{Semi-random maps from a reference ensemble}\label{par:semiRND}
While the polished substrate surface can be approximated as statistically random, the coating process introduces an additional deterministic contribution. 
Because of the large mirror diameter and the several-micrometer thickness of the high-reflective coating stack, the deposited layers are not perfectly uniform at the nanometer level. 
This results in low-spatial-frequency distortions of the reflected wavefront. 
In particular, the surface profile typically remains relatively flat in the central region and progressively decreases toward the edge~\cite{jerome2013Coating}.
This coating-induced component is primarily machine-dependent and therefore expected to be similar for mirrors coated during the same production run. 
Moreover, due to the planetary motion of the mirror during deposition, this contribution can be reasonably approximated as radially symmetric.
Figure~\ref{fig:EMs_coated_zoom} shows the radial cross-sections and radial averages of the AdVirgo+ EM02 and EM04 mirrors, which were coated simultaneously. 
The close agreement between their radially averaged profiles highlights the presence of a common low-order deterministic component superimposed on smaller-scale surface variations. 
This shared behavior motivates the separation between deterministic low-order structure and higher-frequency residuals in the statistical modeling.

Based on a characterization of the reference ensemble, we propose a method to generate semi-random mirror maps that yield controlled surface realizations consistent with the measured distribution of Zernike coefficients. 
Each original surface $\rm{M}(r,\theta)$ is decomposed over a circular aperture using Zernike polynomials,
from which the ensemble mean $\langle c_{nm} \rangle$ and standard deviation $\sigma_{nm}$ are evaluated for each mode.
A new low-spatial-frequency realization is then constructed by defining modified coefficients $c_{nm}^{(\mathrm{new})}$ according to controlled statistical prescriptions. 
Based on the radial symmetry induced by the coating process and confirmed by the manufacturer (LMA), the axisymmetric Zernike modes ($m=0$) up to radial order $n = 0,\dots,14$ are preserved. 
These modes capture the dominant low-spatial-frequency figure components, including the coating-induced radial profile observed in the reference mirrors. 
\begin{equation}
c_{n0}^{(\mathrm{new})} = \langle c_{n0} \rangle,
\qquad n = 0,\dots,14.
\end{equation}
All remaining coefficients (with $m \neq 0$) are sampled from a truncated Gaussian distribution:
\begin{equation}
|c_{nm}^{(\mathrm{new})} - \langle c_{nm} \rangle|
\le k\,\sigma_{nm},
\end{equation}
where $k$ defines the truncation level and prevents unrealistically large deviations from the measured statistical spread.
The truncated Zernike surface is then reconstructed as a linear combination of the new semi-random coefficients.
To retain realistic high-spatial-frequency structure, the high-frequency component extracted from an independent FFT-based map is added: $
\rm{M}_{\mathrm{semiRND\_mix}}(r,\theta)=
Z_{\mathrm{trunc}}(r,\theta)+\rm{M}_{\mathrm{HF}}(r,\theta),
$
where $M_{\mathrm{HF}}$ is obtained by subtracting the Zernike reconstruction of the FFT map from the full FFT surface.

\begin{figure} [h]
    \centering
    \includegraphics[trim=1cm 2mm 1.5cm 1cm clip, width=0.7\textwidth]{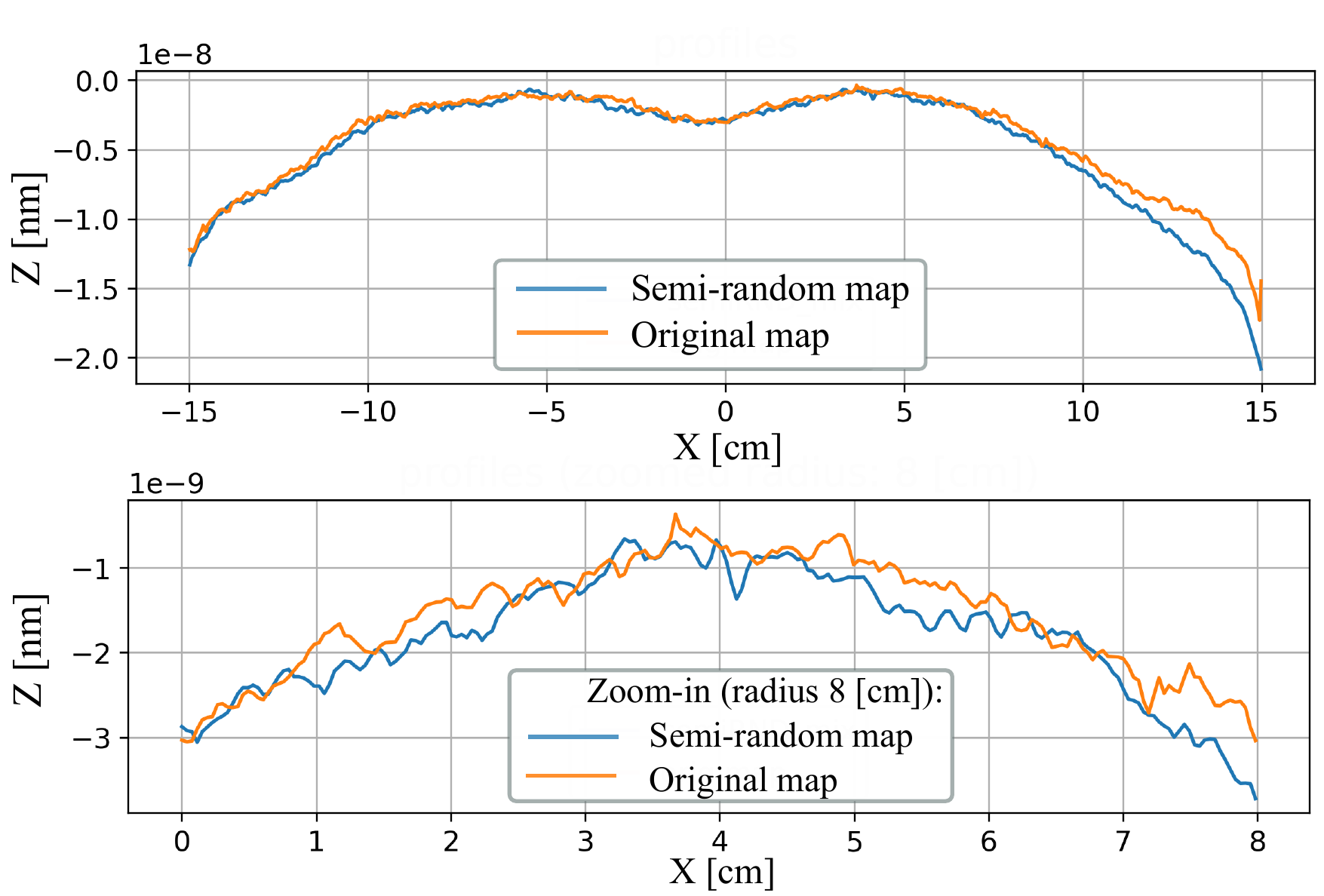}
    \caption{
Comparison of the central surface profiles extracted from the original map (orange) and a semi-random mixed map (blue).
The upper panel shows the full diameter profile, while the lower panel presents a zoom within 8\,cm radius.
The agreement at low spatial frequencies confirms preservation of large-scale figure terms, whereas deviations at smaller scales reflect the stochastic reconstruction of higher-order components.
}.
    \label{fig:semiRND_zoom}
\end{figure}

This procedure ensures consistency with the reference ensemble at low spatial orders while creating physically realistic high-frequency features. We provide an example using EM02 and EM04 mirrors from AdVirgo+, which are selected as the reference set, as they were coated simultaneously and therefore share similar deterministic surface components. The resulting surface realization is shown in Fig.~\ref{fig:semiRND_zoom}.
\subsection{Virtual Mirror Maps of Arbitrary Diameters}
\label{5cm_diam}
Surface measurements of mirrors from previous gravitational wave interferometers can be used to inform the design of future facilities requiring largerdiameter optics; therefore, a strategy to adjust the VMM diameter relative to the source data is necessary.

To generate ET virtual mirror maps, we rescale the data from the AdVirgo+ mirror measurements, whose diameter we call $\text{d}_{\text{V}}$, to the target ET diameter, 
$\mathrm{d_E}$. We therefore define a scaling factor $s$ as:
\begin{equation}
    s = \frac{\mathrm{d_E} }{\mathrm{d_V}}\,.
    \label{eq:scale}
\end{equation}
While phase maps should ideally span the full physical diameter of the mirror, the present study is limited to the region for which experimental surface measurements are available. Accordingly, the virtual mirror maps are constructed with a matching effective diameter.

This scaling factor is applied to the physical size of the numerical VMM data, either by simply rescaling the overall map dimensions, $\mathrm{L_{x,y} }\rightarrow s\mathrm{L_{x,y}}$, or by also adjusting the number of pixels, $\mathrm{N_{x,y}} \rightarrow s\mathrm{N_{x,y}}$, to preserve the spatial resolution.


\begin{figure}[htb]
    \centering
\includegraphics[width=0.9\linewidth]{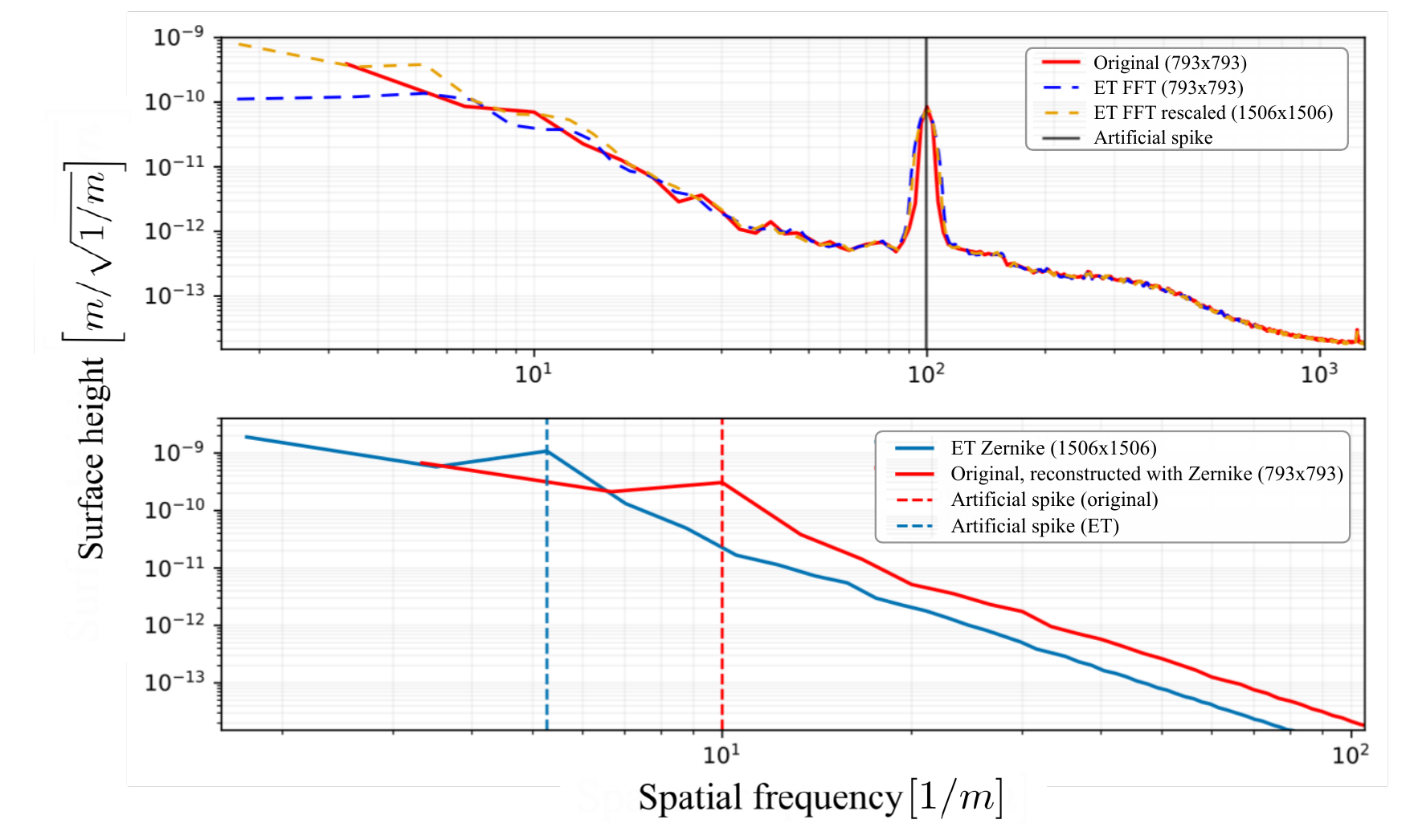}
    \caption{Top panel: comparison between the original map ASD and the two ET cases obtained with the FFT based virtual map method. When the scaling factor is applied also to the number of sampling points, the spatial frequency content of the map is modified. The panel also highlights an artificial spike introduced to confirm that FFT virtual maps do not shift the ASD, they only generate additional points through interpolation in the low spatial frequency region.
    Bottom panel: comparison between the Zernike reconstruction of the original map, after the addition of an artificial spike, and the corresponding Einstein Telescope (ET) Zernike virtual map. After rescaling to the larger ET diameter, the ET Zernike map exhibits a shift of the ASD toward lower spatial frequencies.} 
    \label{fig:scaling}
\end{figure}

Figure~\ref{fig:scaling} emphasizes the different responses of the two approaches. An artificial spectral spike, introduced solely for illustrative purposes, is added to each ASD at $10~\mathrm{m^{-1}}$ (Zernike) and $10^{2}~\mathrm{m^{-1}}$ (FFT).
The upper plot shows the process applied during the FFT VMM generation procedure,
 demonstrating that the rescaling process does not introduce an offset into the spectrum, but the scaling method chosen does impact the range of frequencies, particularly the high-frequency cut-off.
We aim to produce maps with a flatness comparable to that of the original data, as we expect future ET mirrors to be manufactured with similar techniques and therefore to exhibit a similar level of surface flatness. For this reason, we choose to scale also the number of pixels using the same scale factor.

The lower plot depicts the equivalent result when the Zernike VMM generation method is instead used. 
For extended Zernike maps, we first create a larger output grid and then reconstruct the surface by applying the selected $c_{n,m}$ coefficients to it. When these low-order Zernike distortions are expanded to a larger diameter, the surface features are stretched accordingly, which shifts the entire spectral content toward lower spatial frequencies. 
In contrast, for the FFT-\textit{extended} maps, the scale factor is applied to enlarge the 2D mirror map such that the PSD values at the newly generated points are obtained by interpolating the original PSD. This interpolation extends the ASD toward lower spatial frequencies without altering the underlying spectral content. 

\begin{figure} [htb]
    \centering
\includegraphics[width=0.9\textwidth]{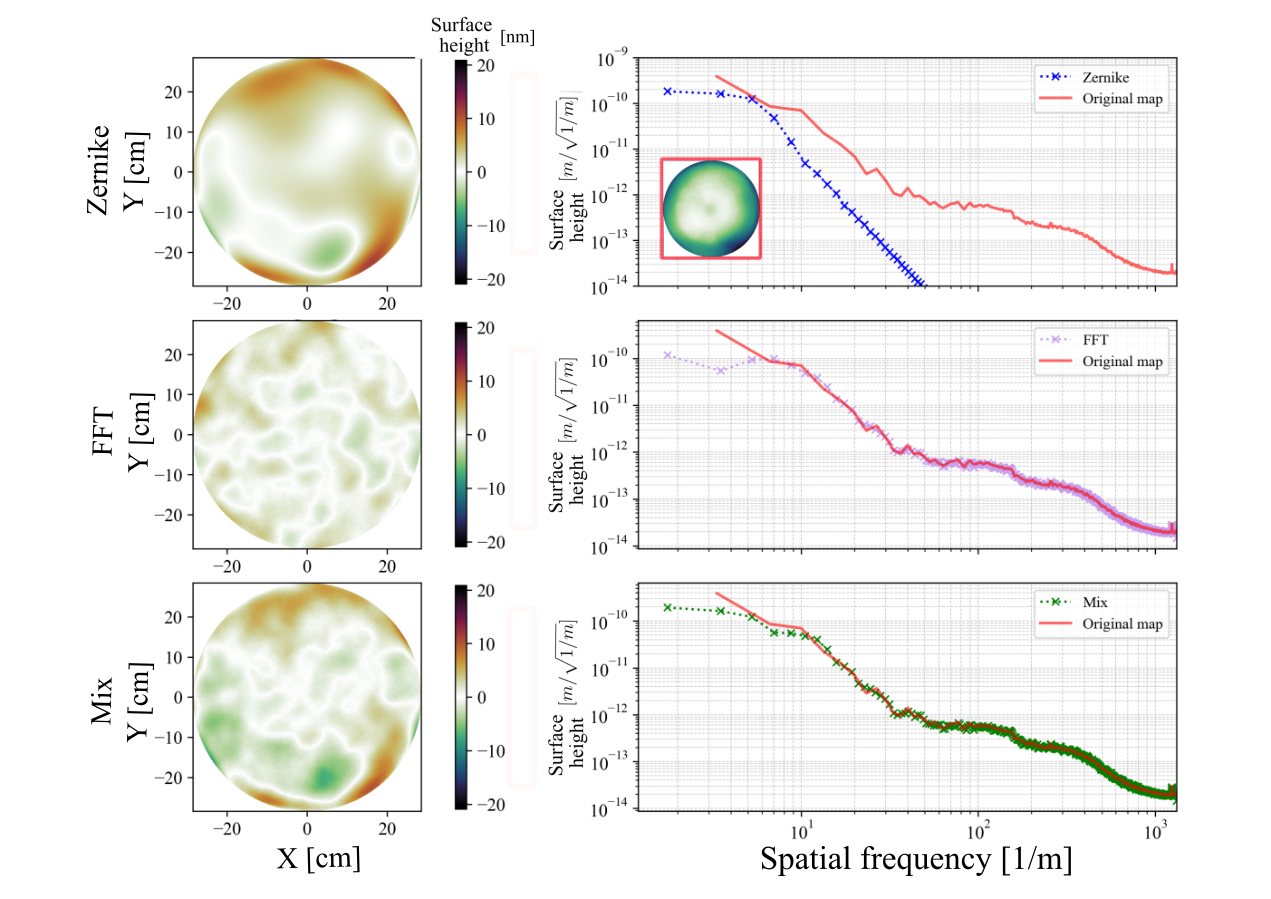}
    \caption{Comparison between the ET virtual mirror maps (left panels) and their corresponding ASDs (right panels) compared to the original map case. In the Zernike ASD plot, the original AdVirgo+ mirror map is shown as a reference in the inset at the top left, highlighted by a red frame.}
    \label{fig:et_orig_psd}
\end{figure}

A comparison between the ET virtual mirror maps generated with the different approaches and their corresponding ASDs is shown in Figure~\ref{fig:et_orig_psd}, illustrating the impact of the chosen reconstruction method on the maps spectral content.

\section{Preprocessing of Map Data for Optical Simulations}\label{sec:map_preparation}
The measurement process used to obtain the mirror phase maps of the physical optical surfaces does not take into account that actuators can position and deform the mirrors to compensate for some simple imperfections once they are installed in the detector, particularly the lowest order surface defects: \textit{piston}, \textit{tilt}, and \textit{curvature}. 
These effects can be studied independently in optical simulations, therefore to understand the optical performance of a surface, they should be removed from the map data. 
We therefore compare two \textit{preprocessing} strategies to determine which best prepares mirror surface maps that are suitable for optical models of gravitational-wave detectors, taking into account the laser beam shape and the mirror regions most critical for the interferometer performance. 


\paragraph{The Zernike approach} determines the coefficients $c_{n,m}$ for each relevant degree of freedom as described in section~\ref{subsec:framework}, enabling a direct identification and quantification of optical aberrations on the mirror surface. This is a well-established approach in the field \cite{Evans1995_VisualizationZernike, Mahajan1995_ZernikeOptAb}. To remove piston, tilt, and curvature from a raw mirror map, we compute the Zernike basis up to order $n=2$ and isolate the corresponding coefficients. A map containing only these components is then subtracted from the original, effectively filtering out these imperfections while preserving physical features like astigmatism.


\paragraph{The Hermite-Gauss method} instead employs a strategy by which the calculation of the piston, tilt and curvature is weighted by the Gaussian intensity profile of the incident optical field. 
The method, utilized by the \Finesse software package \cite{brown_2025_12662017}, has the advantage of naturally emphasizing the regions of the mirror that have the greatest impact on the optical field, providing a more suitable representation of the effective surface seen by the laser beam. 
This approach computes overlap integrals of the ideal Gaussian incoming beam and the relevant higher-order Hermite-Gauss mode in the reflected beam with the map data to extract the values of piston, tilt and curvature. With $w$ the spot size of the targeted laser beam, $\delta_x = x_1 - x_0$ and $\delta_y = y_1 - y_0$ the grid spacing of the map data and $w' = w/\sqrt{2}$ we get:
\begin{align}
c_{\rm piston}
  &= \frac{\delta_x\,\delta_y}{\pi w'^2}
     \sum_{i,j} \mathrm{M}_{ij}\,
     \exp\!\left(-\frac{x_i^2 + y_j^2}{w'^2}\right),
\\
c_{\rm x\text{-}tilt}
  &= \frac{2\,\delta_x\,\delta_y}{\pi w'^4}
     \sum_{i,j} \mathrm{M}_{ij}\, x_i\,
     \exp\!\left(-\frac{x_i^2 + y_j^2}{w'^2}\right),
\\
c_{\rm x\text{-}curv}
  &= \frac{\delta_x\,\delta_y}{2\pi w'^4}
     \sum_{i,j} \mathrm{M}_{ij}
     \left[\left(\frac{2x_i}{w'}\right)^2 - 2\right]
     \exp\!\left(-\frac{x_i^2 + y_j^2}{w'^2}\right).
\end{align}
and similarly the coefficients for $c_{\rm y\text{-}tilt}$ and $c_{\rm y\text{-}curv}$. We can then compute the corrected map as:
\begin{equation}
\mathrm{M}_{\rm corrected}(x, y)
  = \mathrm{M}(x, y)
  - c_{\rm piston}
  - c_{\rm x\text{-}tilt}\,x
  - c_{\rm y\text{-}tilt}\,y
  - c_{\rm x\text{-}curv}\,x^2
  - c_{\rm y\text{-}curv}\,y^2.
\end{equation}

\begin{figure} [htb]
    \centering
    \includegraphics[width=0.9\textwidth]{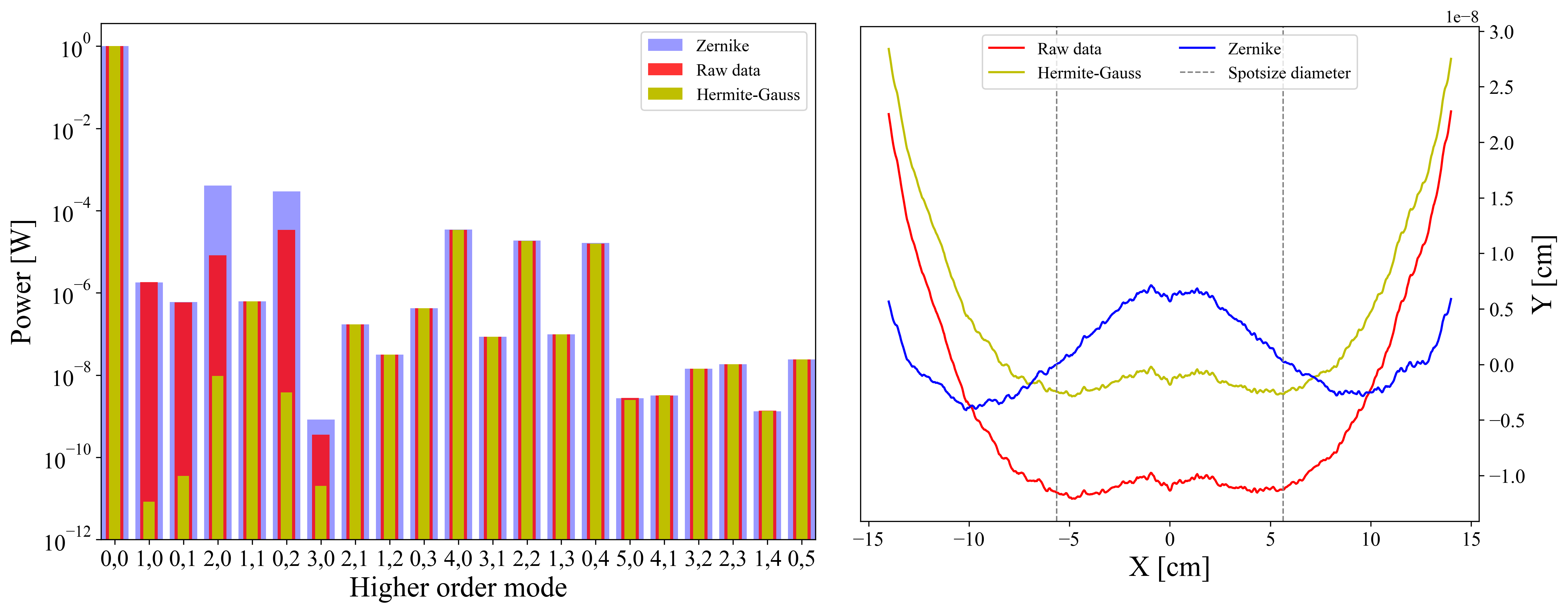}
    \caption{
Left: Distribution of power into higher-order modes (HOMs) calculated after removing tilt and curvature from the original map, using two different methods based on Zernike and Hermite-Gauss polynomials. The raw (not preprocessed) data is included for reference. \\
Right: cross-sectional surface profiles of the same maps; grey vertical lines highlight the incident beam spot diameter on the mirror.}
\label{fig:prt_paper}
\end{figure}

The subtraction of the quadratic terms removes the average curvature (defocus) and the astigmatism component aligned with the x and y axes. However, the full astigmatic content is not removed (e.g. rotated astigmatism terms proportional to 
$xy$), in order to remain consistent with the way mirror maps are treated in \Finesse.
Figure~\ref{fig:prt_paper} shows the effect of each preprocessing method by computing the optical field directly reflected from the resulting surface (for tilt and radius of curvature removal; piston removal is not necessary in the case of a single mirror, as the absolute phase of the reflected beam does not matter because there is no resonance condition to satisfy). 
The Hermite-Gauss method significantly reduces the power content in several higher-order modes (HOMs) associated with the targeted distortions, while the Zernike approach does not significantly reduce the HOM  content, and in some cases even worsens it.  
Furthermore, we have verified that the sequence in which piston, tilt and curvature are removed does not have a significant impact on the optical performance of the surface. 

\section{Validation of Virtual Mirror Maps for Einstein Telescope design}\label{sec:results}

We choose to validate the VMM generation methods above based on their optical performance, as this directly targets the mirrors' intended application in gravitational-wave interferometry. 
We adopt a statistical approach and evaluate a set of figures of merit over ensembles of 1000 maps per method, for both the AdVirgo+ and ET cases. 

Each generation method is used to create a set  of VMMs with randomized surface structures but comparable power spectral densities and overall flatness. 
To make the comparison meaningful, each VMM next undergoes the same preprocessing steps applied to the measured mirror data. 
The processed maps 
are then sequentially applied to a \Finesse-based optical simulation of a single mirror under direct reflection, returning the total optical loss from the fundamental mode and scattering into higher-order modes. 
The statistical spread of results reflects the variability introduced by the randomized surface fluctuations, while the average behavior reflects the expected performance of the corresponding mirror population.

Two test optical configurations are considered. 
We first use parameters based on the AdVirgo+ interferometer design to confirm that the methods can be used to reasonably replicate the optical performance of measured AdVirgo+ mirrors. 
Then, we extend the approach to Einstein Telescope High-Frequency (ET-HF) design parameters, allowing us to verify that the behavior of the algorithms remains consistent when extrapolated to larger mirror sizes.

In both cases, the optical setup is defined as follows: a 1\,W Gaussian beam, with properties matched to the curvature of the mirror,  propagates along the \(z\)-axis (the \(x\)-axis is aligned to the horizontal plane of the interferometer). 
The beam hits the mirror, which has reflectivity $\mathrm{R}=1$, at normal incidence.
The field content is computed immediately upon reflection from the mirror surface with no further propagation.
The key optical and geometrical parameters used in the two test configurations, referring exclusively to the End Mirrors, are listed in Table~\ref{tab:et_virg_optSpec} and are based on their respective design documents~\cite{ET_TDS_17461,virgo_design}.






\begin{table}[h!] 
\centering

\begin{tabular}{
  l
  r
  @{\hspace{6mm}}
  c
  @{\hspace{6mm}}
  c
}
\toprule
&& \textbf{ET-HF} & 
\textbf{AdVirgo} \\
\midrule
\midrule
\multicolumn{4}{l}{\textbf{Beam}}\\
\midrule

$\lambda$ & [nm] & 1064 & 1064 \\
$w$       & [mm] & 120.47 & 56.38  \\
$w_0$     & [mm] & 14.15  & 10.03  \\

\midrule
\midrule
\multicolumn{4}{l}{\textbf{End Mirror}}\\
\midrule

$\mathrm{R_C}$  & [m]  & 5070.0  & 1692.2 \\
$\diameter$     & [m]  & 0.62    & 0.35   \\

\bottomrule
\end{tabular}
    \caption{Beam and mirror specifications used in the simulation setups for the ET-HF and AdVirgo+ scenarios. Symbols: $\lambda$, laser wavelength; $w$, Gaussian beam radius at the optic; $w_{0}$, beam waist radius; $\mathrm{R_{C}}$, mirror radius of curvature; $\diameter$, optic (clear aperture) diameter.}
	\label{tab:et_virg_optSpec}
\end{table}

\subsection{Validation against AdVirgo+ measured data}
We first assess whether the generated VMMs reproduce the surface characteristics and corresponding optical losses achieved using measured AdVirgo+ mirror surface data. 

\begin{figure}[h!]
  \centering
\includegraphics[width=0.8\textwidth]{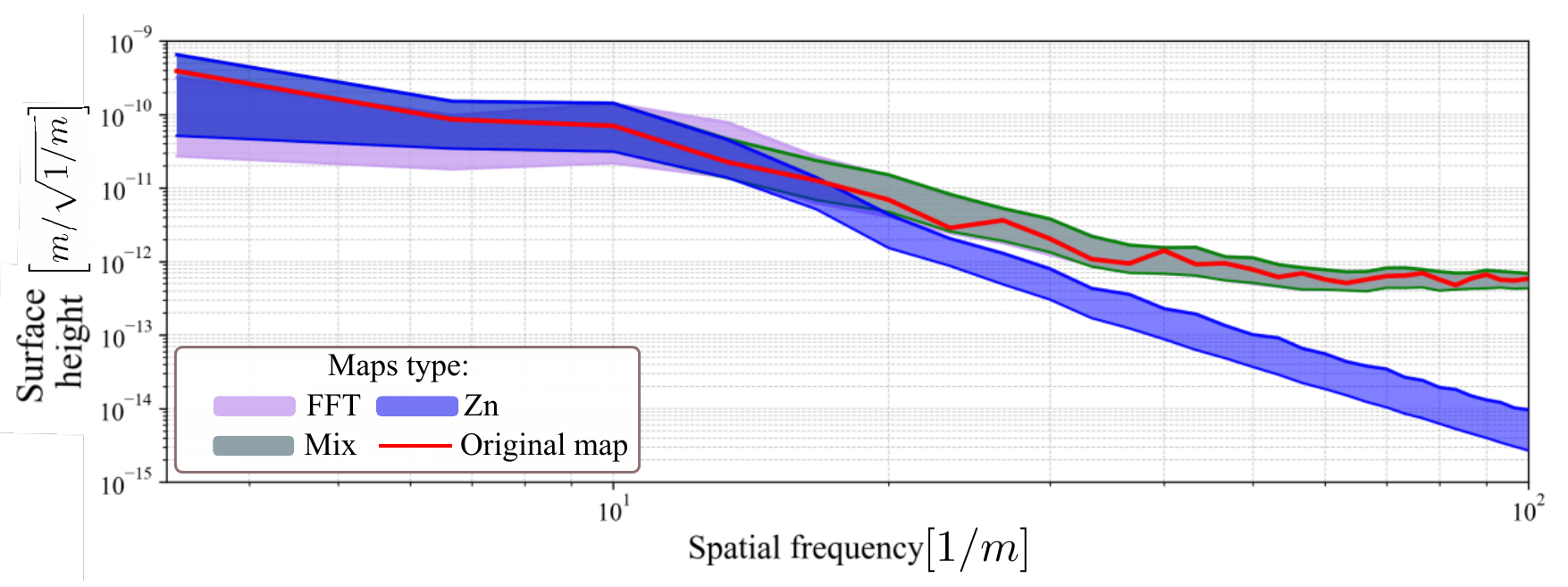}
  \caption{Range of ASDs produced by AdVirgo+ VMMs generated with the Zernike, FFT, and mixed methods (1000 maps per method),  together with the ASD of the original pre-processed AdVirgo+ map. While the Zernike maps underestimate the high frequency tail of the spectrum, the FFT and mixed methods reproduce the original ASD more accurately across the full spatial frequency range.}
  \label{fig:AdVirgo_psd1}
\end{figure}

Figure~\ref{fig:AdVirgo_psd1} compares the 1D ASD of an AdVirgo+reference map with the ensembles of VMMs generated using the Zernike, FFT, and mixed approaches. 
The shaded regions indicate the minimum and maximum envelope of ASDs found across 1000 maps, evaluated at each spatial-frequency bin. 
As anticipated, the Zernike maps reproduce the low-spatial-frequency region but underestimate the high-frequency tail. 
The FFT maps show slightly lower amplitudes at low frequencies, with a wider range of surface heights than the Zernike case, but accurately follow the reference at higher frequencies. Finally, as expected, the mixed method provides the most faithful reproduction of the full spectrum.

\begin{figure}[h!]
  \centering
\includegraphics[width=0.65\textwidth]{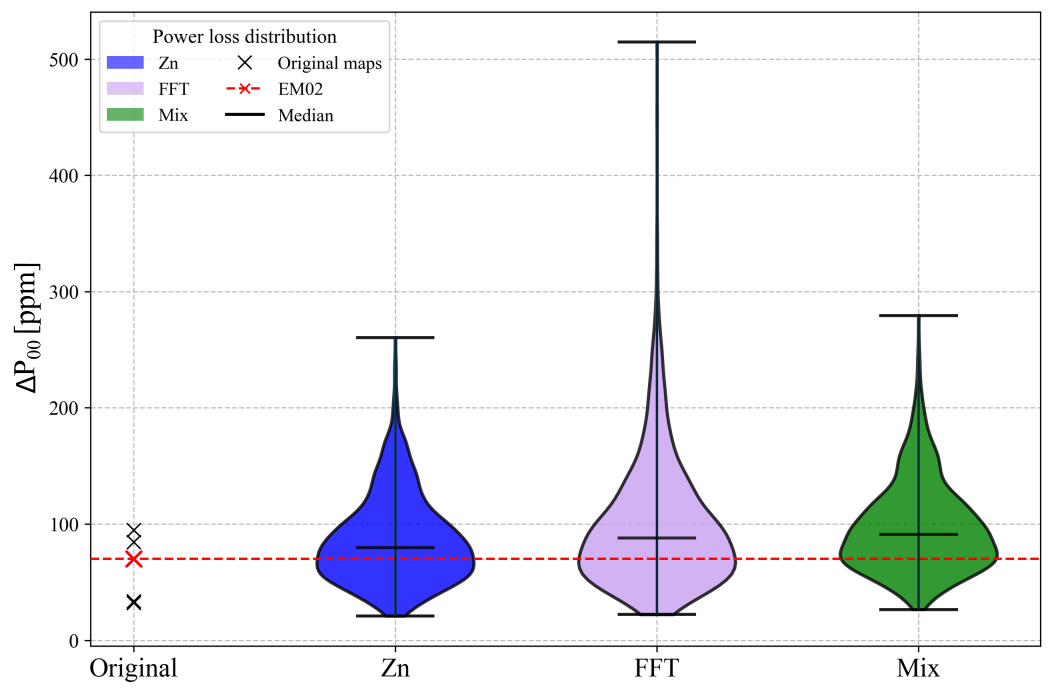}
    \caption{
        Violin plot showing the distribution of optical power losses, in the fundamental mode, obtained for AdVirgo virtual mirror maps using different generation methods. The categories include the original maps (Originals), Zernike based reconstructions (Zn), FFT based reconstructions (FFT), and mixed reconstructions (Mix). The horizontal lines indicate the median values for each dataset, while the reference level corresponding to the EM02 phase map is shown for comparison, since this is the original AdVirgo+ map used as starter for all our VMMs. The data set for each violin plot contains one thousand maps each. 
    }
    \label{fig:powloss_violin}
\end{figure}

Figure~\ref{fig:powloss_violin} reports the power losses from the fundamental mode for each method, compared to six measured AdVirgo+ maps (four corresponding to the mirrors installed in the interferometer and two spare mirrors). The loss is defined as:
\begin{equation}
    \Delta \text{P}_{00} = \text{P}_{00}^{\mathrm{Inj}} - \text{P}_{00}^{\mathrm{Refl}},
\end{equation}
where $\Delta \text{P}_{00}$ is the variation in the carrier power, defined as the difference between the total injected power, $\text{P}_{00}^{\mathrm{Inj}}$, and the power contained in the reflected carrier field, $\text{P}_{00}^{\mathrm{Refl}}$. Note that this definition of loss is different to other methods used in literature, such as for example the rountrip loss from the total field in a cavity. The red cross marks the original mirror map used to generate all the VMMs. Zernike-based maps exhibit systematically lower losses, consistent with the absence of fine scale flatness, while the FFT and mixed methods produce higher loss values.
The thin tail towards high loss in the violin plots arise from the poorest realizations among each set of VMMs: our ensembles sample a broad range of possible surfaces that match the provided 1D ASD and/or Zernike  coefficients, more than the limited set provided by the manufacturers, and include maps that represent mirrors that would normally be discarded.

\begin{figure}[h!]
  \centering
  \includegraphics[width=\textwidth]{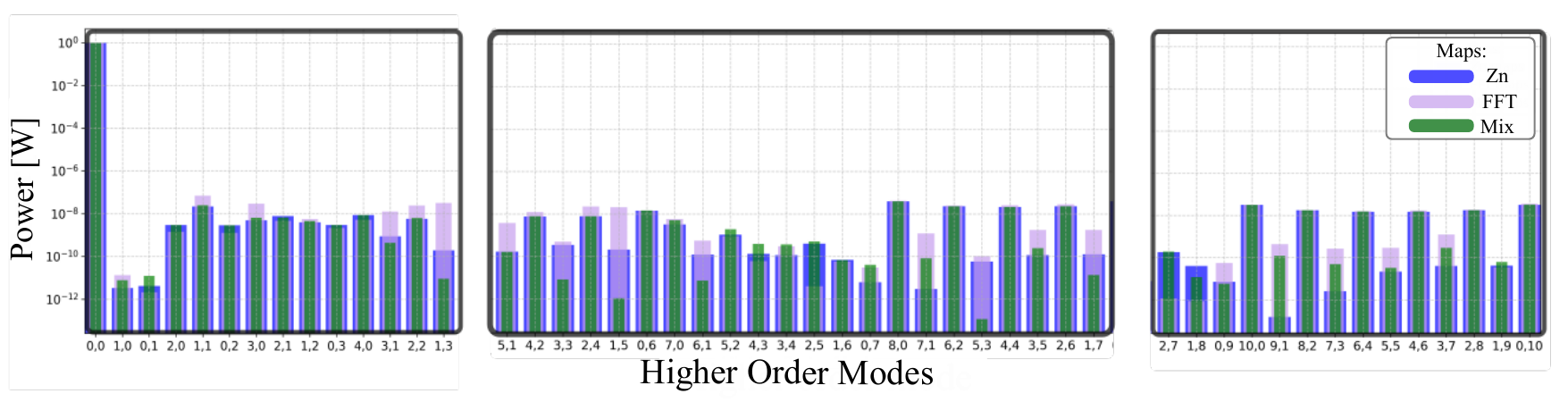}
  \caption{Selected sections of the Higher Order Modes power content in the AdVirgo+ virtual maps. The compact display highlights how the surface distortions introduced by the different reconstruction methods scatter power not only into the lowest HOMs but also into higher order ones. For each method, the plotted values represent the average power content for each mode, computed over the full ensemble of maps.} 
  \label{fig:virgo_hist}
\end{figure}

 The excess loss of these poor realizations manifests as power scattered into HOMs.
Figure~\ref{fig:virgo_hist} shows the average HOM power content for each VMM method from an ensemble of one thousand maps, showing the distribution of optical power up to maximum Hermite-Gauss order of 10. 
The lowest HOMs typically carry less power with respect to the others due to the preprocessing applied (see section~\ref{sec:map_preparation}). Note that other higher order contributions, although not explicitly shown, are also present.

To quantify the surface differences across the different reconstruction methods, we evaluate the RMS flatness of the surface maps, reported as the mean $\pm$ standard deviation for each set of virtual mirror maps. Over the full usable aperture, corresponding to a radius of approximately $14\,\mathrm{cm}$, the original Virgo maps exhibit an average RMS flatness of $5.58 \pm 1.77\,\mathrm{nm}$. All virtual mirror maps yield lower RMS values, with $4.16 \pm 0.86\,\mathrm{nm}$ for the Zernike-based method, $2.48 \pm 0.74\,\mathrm{nm}$ for the FFT-based method, and $4.16 \pm 0.85\,\mathrm{nm}$ for the mixed approach.

When restricting the calculation to a $15\,\mathrm{cm}$ diameter region around the map center, corresponding to the area most relevant for the incident laser beam, the RMS flatness decreases for all reconstruction methods. In this central region, the original Virgo maps reach an average RMS of $0.46 \pm 0.11\,\mathrm{nm}$, while the virtual mirror maps yield $0.89 \pm 0.23\,\mathrm{nm}$ for the Zernike-based method, $0.96 \pm 0.27\,\mathrm{nm}$ for the FFT-based method, and $0.94 \pm 0.22\,\mathrm{nm}$ for the mixed approach. When the analysis is restricted to the central region, the RMS values of the different virtual mirror maps become comparable, reflecting the reduced contribution of large-scale surface features.
These values are consistent with the ASDs plots 
showed before in Figure~\ref{fig:AdVirgo_psd1}.

\begin{figure}[h!]
  \centering
  \includegraphics[width=\textwidth]{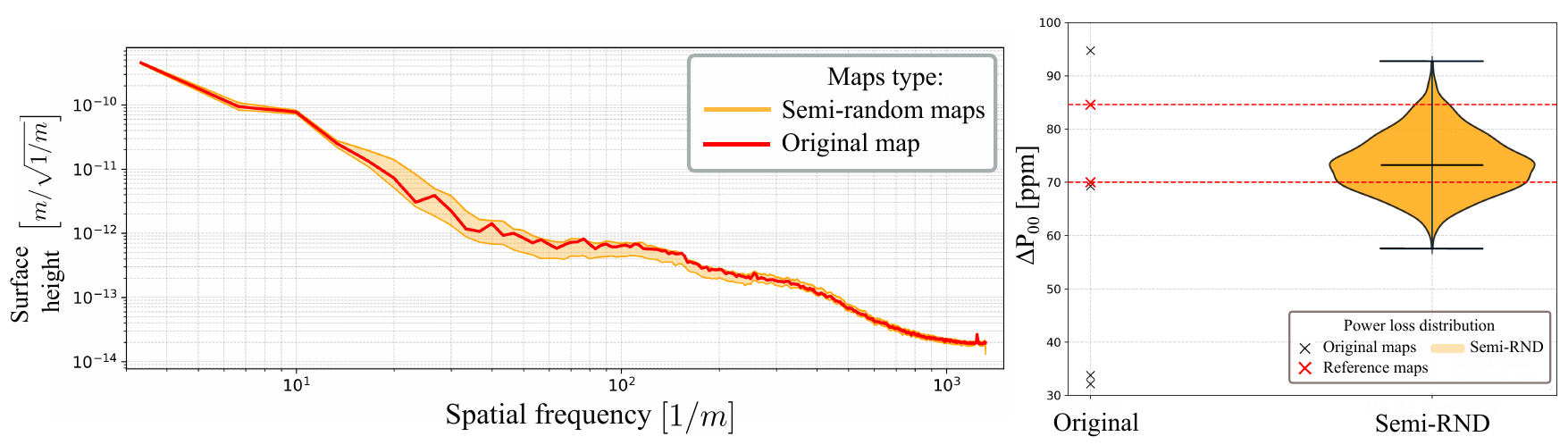}
  \caption{Comparison between the original mirror maps and the semi-random (Semi-RND) maps optical performances.
Left: Radially averaged 1D amplitude spectral density of the surface height as a function of spatial frequency. The Semi-RND maps reproduce the spectral behaviour of the reference map (AdVirgo+ EM02) across the full spatial frequency range. The Semi-RND ensemble is generated from a combination of the two original maps (EM02 and EM04).
Right: Distribution of the fundamental mode power loss $\Delta P_{00}$. The violin plot shows the Semi-RND ensemble, while crosses mark the values obtained for the original maps (EM02 and EM04). The Semi-RND maps produce losses within the range defined by the reference mirrors.}
  \label{fig:semiRND_psd_pow}
\end{figure}

In addition to the fully randomized virtual mirror maps described above, we also consider the semi-random mix generation scheme (as introduced in paragraph~\ref{par:semiRND}), in which the dominant axisymmetric Zernike modes are preserved from the reference ensemble while the remaining modes are randomized within the measured statistical spread. This approach retains the characteristic radial surface profile associated with the coating process while allowing stochastic variability in the higher-order components.\\
In Figure~\ref{fig:semiRND_psd_pow} the resulting one-dimensional ASD and the corresponding carrier power losses obtained from ensembles of semi-random maps are shown. The generated surfaces reproduce the spectral behaviour of the reference mirrors and yield comparable optical losses. The distributions of higher-order mode content and the RMS flatness are also consistent with those of the original mirror maps.

\subsection{Virtual Mirror Maps applied for the ET design case}
\begin{figure}[htb]
  \centering  \includegraphics[width=0.8\textwidth]{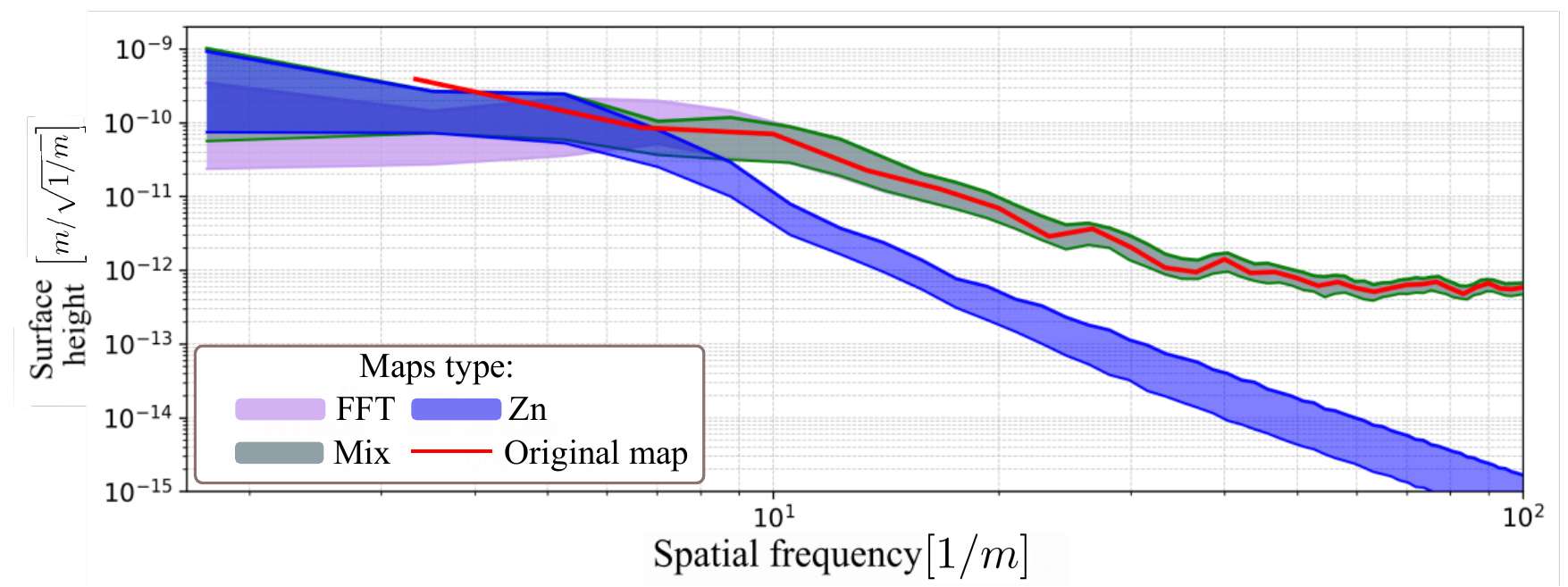}
  \caption{One-dimensional Amplitude Spectral Densities (ASDs) of the ET-scaled virtual mirror maps generated using the Zernike, FFT, and mixed methods. 
    Shaded regions represent the minimum-maximum envelopes over one thousand realizations for each method. 
    The ASD of the original AdVirgo+ reference map is shown for comparison. 
    The Zernike maps underestimate the high frequency tail, the FFT maps reproduce the high frequency behaviour but deviate at low frequencies, and the mixed method provides the best agreement across the full spatial frequency range.}
  \label{fig:et_psd1}
\end{figure}

\begin{figure}[htb]
  \centering
  \includegraphics[width=0.65\textwidth]{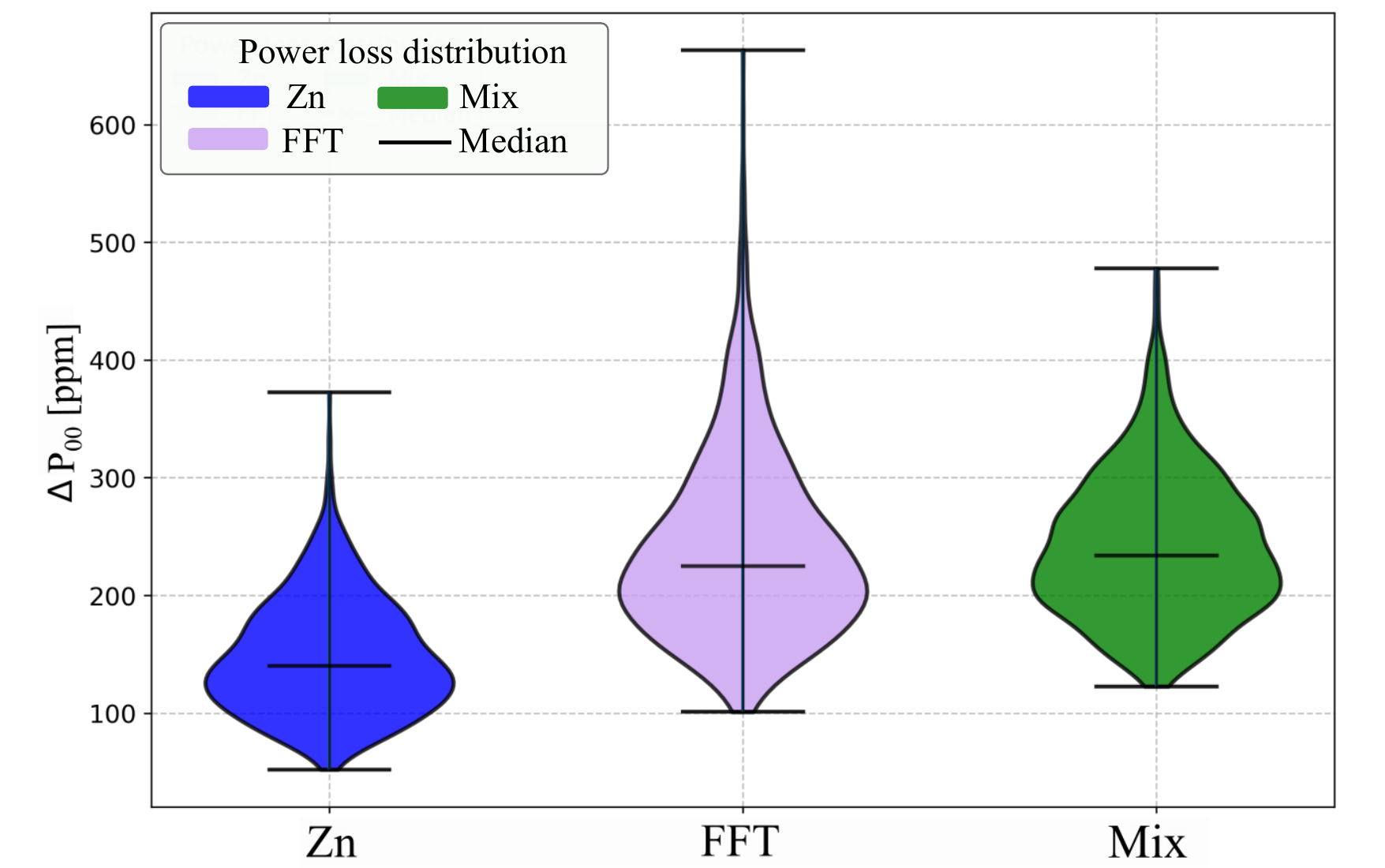}
    \caption{
    Violin plots of the carrier-mode power loss obtained for the ET-HF simulations using Zernike, FFT, and mixed virtual mirror maps. 
    Each distribution contains one thousand realizations. 
    Zernike maps show systematically lower losses due to their smoother profiles, while FFT maps produce higher losses and a broader statistical spread. 
    The mixed method yields realistic and statistically stable loss values, confirming its suitability for ET scale synthetic surfaces.
    }
    \label{fig:ET_loss}
\end{figure}

\begin{figure}[htb]
  \centering
  \includegraphics[width=\textwidth]{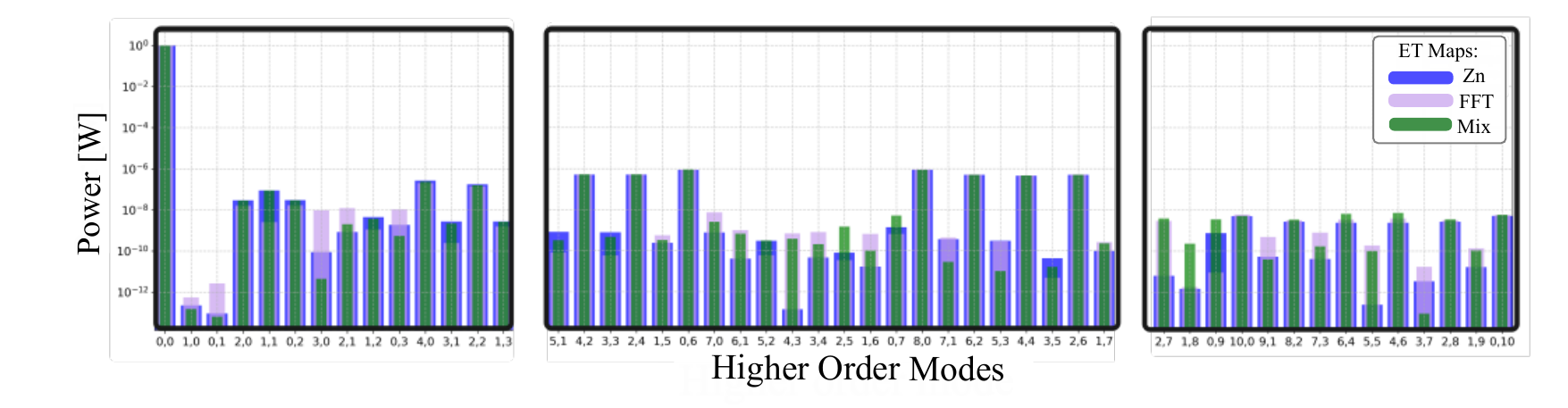}
  \caption{Distribution of the power scattered into Hermite–Gauss higher-order modes (up to order 10) for the ET-scaled virtual mirror maps produced with the Zernike, FFT, and mixed reconstruction methods. For each method, the plotted values represent the average power in each mode, computed over the ensemble of one thousand maps.}
  \label{fig:et_hom}
\end{figure}

We now assess the performance of the VMM generation methods in the context of the Einstein Telescope High-Frequency (ET-HF) design. This configuration requires significantly larger mirror diameters and beam radii (see Table~\ref{tab:et_virg_optSpec}), and serves as stringent test of the methods. 
We again generate 1000 VMMs per method, now rescaled as described in Section~\ref{5cm_diam}. \\
Figure~\ref{fig:et_psd1} reports the resulting 1D ASDs, shown as minimum-maximum envelopes across each ensemble, together with the ASD of the original AdVirgo+ reference map.
The performances of the Zernike, FFT and mixed methods are consistent with that found for the Virgo case; now with additional content at lower frequencies due to the extension in diameter. 
The mixed method continues to provide the best global agreement with the reference ASD across the full frequency range, confirming that this strategy remains effective even after extrapolation to ET-scale diameters.
A similarly consistent trend is found when analyzing the optical losses, as summarized in Figure~\ref{fig:ET_loss}. 
The mixed-mode ensemble again occupies an intermediate regime between the Zernike and FFT VMMs. 
Figure~\ref{fig:et_hom} shows the average power scattered into HOMs, up to Hermite-Gauss order 10. 
No particular VMM type shows a distinctly better performance than another from this perspective, however 
all HOM contributions remain well below the $1\,\mathrm{ppm}$ level. 

To further characterize the surface statistics of the ET-scaled virtual mirror maps, we evaluate the RMS flatness computed from the surface maps over the full usable aperture and within a $30\,\mathrm{cm}$ diameter region around the map center. Over the full aperture, the Zernike-based maps exhibit an average RMS flatness of $3.74 \pm 0.70\,\mathrm{nm}$, while the FFT-based maps reach lower values of $1.78 \pm 0.32\,\mathrm{nm}$. The mixed maps show RMS values comparable to the Zernike case, with $3.82 \pm 0.68\,\mathrm{nm}$. Restricting the analysis to a $30\,\mathrm{cm}$ diameter central region, the RMS flatness decreases for all reconstruction methods. In this case, the Zernike maps yield $0.96 \pm 0.22\,\mathrm{nm}$, the FFT-based maps $1.28 \pm 0.20\,\mathrm{nm}$, and the mixed maps $1.26 \pm 0.19\,\mathrm{nm}$. 
When restricting the calculation to the central 15\,cm, all methods yield reduced RMS values, as expected, with the FFT and mixed maps showing nearly identical central flatness.

Overall, the ET VMMs closely `mirror' the behavior observed in the AdVirgo+ case, demonstrating that the methods remain stable and qualitatively meaningful when applied to ET-sized mirrors. These results confirm that, based on our chosen figures of merit, the mixed method is preferable for generating Virtual Mirror Maps for the Einstein Telescope, combining realistic surface properties with consistent performance across spatial frequencies and optical figures of merit.


\section{Conclusions}

We have developed a framework for generating `virtual mirror maps'.  
These are synthetic mirrors that reproduce the spatial properties of real optical surfaces while introducing controlled randomization.
This enables mirror surface specifications to be developed and tested in optical models during the design of future interferometers, such as the proposed Einstein Telescope.
Our VMM generation methods may be tuned to target user-specified figures of merit such as Power Spectral Density (PSD) and total optical loss. 
The resulting virtual maps remain statistically consistent with the reference data for any chosen mirror diameter.

We validated the framework using AdVirgo+ mirror data as a relevant reference and realistic baseline for the optical specifications of ET, qualifying our results based on the resulting optical performance. 
virtual maps intended to replicate AdVirgo+ data produced optical losses that are fully comparable to measured mirror surface data. 
We further showed that the procedure remains stable when extrapolating to ET-scale diameters.

We have shown how the Hermite--Gauss based pre-processing provides a more effective preparation of mirror surface maps for optical simulations. By weighting surface distortions according to the carrier beam intensity, it more efficiently suppresses higher-order mode content within the beam footprint, while the Zernike-based approach does not consistently achieve this. As a result, the Hermite--Gauss method yields a more physically relevant representation of the effective mirror surface seen by the interferometer.
We compared Zernike-based, FFT-based, and hybrid VMM generation methods in detail, finding that the choice of method has a significant impact on the resulting VMM performance. 
The Zernike method is effective when low-spatial-frequency distortions are of interest, while the FFT-based approach preserves high spatial frequency features. 
We find that the mixed method combines the advantages of both approaches and yields the most robust agreement with the reference spatial statistics and optical losses across all spatial frequencies.

This also implies that one must be careful when defining specifications for mirror surfaces (this is equivalent to choosing a VMM generation method), as each type of specification (method) will constrain the important figures of merit for the experiment differently. 

Overall, the algorithms presented here provide a flexible and statistically reliable tool for generating virtual mirror surfaces of arbitrary size, enabling realistic optical simulations for current and future gravitational-wave detectors. 


\section*{Acknowledgments}
We thank the Laboratoire des Matériaux Avancés for many useful discussions, and for providing the AdVirgo+ phase maps used in this study. 
The authors gratefully acknowledge the Italian Istituto Nazionale di Fisica Nucleare (INFN),  
the French Centre National de la Recherche Scientifique (CNRS) and
the Netherlands Organization for Scientific Research (NWO), 
for the construction and operation of the Virgo detector
and the creation and support of the EGO consortium.
The authors also gratefully acknowledge research support from these agencies as well as by 
the Spanish  Agencia Estatal de Investigaci\'on, 
the Consellera d'Innovaci\'o, Universitats, Ci\`encia i Societat Digital de la Generalitat Valenciana and
the CERCA Programme Generalitat de Catalunya, Spain,
the National Science Centre of Poland and the European Union – European Regional Development Fund; Foundation for Polish Science (FNP),
the Hungarian Scientific Research Fund (OTKA),
the French Lyon Institute of Origins (LIO),
the Belgian Fonds de la Recherche Scientifique (FRS-FNRS), 
Actions de Recherche Concertées (ARC) and
Fonds Wetenschappelijk Onderzoek – Vlaanderen (FWO), Belgium,
the European Commission.
The authors gratefully acknowledge the support of the NSF, STFC, INFN, CNRS and Nikhef for provision of computational resources
This publication is part of the project “Smoothing the Optical Bumps in the Road for Future Gravitational-Wave Detectors”, led by A.~C.~Green, with project number VI.Veni.212.047, which is financed by the Dutch Research Council (NWO).
The authors gratefully acknowledge the support of the NSF, STFC, INFN, CNRS and Nikhef for provision of computational resources. 
The authors also thank Prof.~H.~Yamamoto, Dr.~Y.~Guo, and M.~Eisenmann for useful discussions during the Finesse Workshop held at the Nikhef Institute in Amsterdam. We also thank M. van der Kolk, J. Gobeil, and J. Richardson for their valuable support.


\bibliographystyle{unsrtnat}
\bibliography{references}
\end{document}